\documentclass[runningheads]{llncs}
\usepackage[T1]{fontenc}
\usepackage{graphicx,verbatim}
\usepackage{amsmath}
\usepackage{amssymb}
\usepackage{multirow}
\usepackage{booktabs}
\usepackage{marvosym}
\usepackage{url}
\begin{document}
\title{Gabor Primitives for Accelerated Cardiac Cine MRI Reconstruction}
\titlerunning{Gabor Primitives for Accelerated Cardiac Cine MRI}

\author{
Wenqi Huang\inst{1,2} \Letter \and
Veronika Spieker\inst{2,3,4} \and
Nil Stolt-Ansó\inst{1,2,5} \and
Natascha Niessen\inst{2,6} \and
Maik Dannecker\inst{1,2} \and
Sevgi Gokce Kafali\inst{1,2} \and
Sila Kurugol\inst{4} \and
Julia A. Schnabel\inst{2,3,5,7} \and
Daniel Rueckert\inst{1,2,5,8}
}

\authorrunning{W. Huang et al.}

\institute{
Chair for AI in Healthcare and Medicine, Technical University of Munich (TUM) and TUM University Hospital, Munich, Germany \and
School of Computation and Information Technology, Technical University of Munich, Munich, Germany \and
Institute of Machine Learning in Biomedical Imaging, Helmholtz Munich, Neuherberg, Germany \and
Department of Radiology, Boston Children's Hospital and Harvard Medical School, Boston, USA \and
Munich Center for Machine Learning, Technical University of Munich, Munich, Germany \and
GE HealthCare, Munich, Germany \and
School of Biomedical Engineering and Imaging Sciences, King's College London, London, United Kingdom \and
Department of Computing, Imperial College London, London, United Kingdom \\
\email{wenqi.huang@tum.de}
}


\maketitle
\begin{abstract}
Accelerated cardiac cine MRI requires reconstructing
spatiotemporal images from highly undersampled
\textit{k}-space data. 
Implicit neural representations (INRs) enable scan-specific reconstruction without large training datasets, but encode content implicitly in network weights without physically interpretable parameters.
Gaussian primitives provide an explicit and geometrically interpretable alternative, but their spectra are confined near the \textit{k}-space origin, limiting high-frequency representation.
%
We propose Gabor primitives for MRI reconstruction, modulating each Gaussian envelope with a complex exponential to place its spectral support at an arbitrary \textit{k}-space location, enabling efficient representation of both smooth structures and sharp boundaries.
To exploit spatiotemporal redundancy in cardiac cine, we decompose per-primitive temporal variation into a low-rank geometry basis capturing cardiac motion and a signal-intensity basis modeling contrast changes.
%
Experiments on cardiac cine data with Cartesian and radial trajectories show that Gabor primitives consistently outperform compressed sensing, Gaussian primitives, and hash-grid INR baselines, while providing a compact, continuous-resolution representation with physically meaningful parameters.
\footnotemark

\end{abstract}
\footnotetext{Code will be released upon acceptance.}
%
%
%
\section{Introduction}

Cardiac cine MRI requires high spatial and temporal resolution, yet the achievable resolution is constrained by scan duration, which is limited by MR physics and patients' tolerance~\cite{rajiah2023cardiac,wang2021fast}. Undersampling \textit{k}-space is therefore essential, but results in an ill-posed inverse reconstruction problem that requires effective priors or regularization.

Compressed sensing (CS) and low-rank methods exploit sparsity and temporal redundancy~\cite{feng2022golden,otazo2015low}, but their hand-crafted regularizers may struggle with complex, patient-specific motion and fine anatomical detail. Supervised deep learning methods can learn stronger data-driven priors~\cite{hammernik2018learning,qin2018convolutional}, yet require large training datasets and risk hallucination on out-of-distribution cases~\cite{antun2020instabilities,johnson2021robustness}. 
Implicit neural representations (INRs) offer a scan-specific alternative requiring no external data~\cite{feng2025spatiotemporal,huang2023neural,spieker2023iconik}, but encode information implicitly in network weights with no physical interpretability.

In computer vision, 3D Gaussian splatting (3DGS)~\cite{kerbl20233dgs} represents scenes as collections of Gaussian primitives with learnable position, shape, color and opacity. This idea has recently been adapted to MRI reconstruction~\cite{singh2026gaussian,terpstra2026gaussian}. 
However, individual Gaussian primitives are inherently smooth in image space, making sharp features such as tissue boundaries costly to represent. In the frequency domain, their spectra remain centered at the \textit{k}-space origin, as spatial translation introduces only a phase shift but not a frequency shift. Computer vision has addressed this high-frequency limitation using frequency-modulated primitives such as wavelet~\cite{wipes2025} and Gabor splatting~\cite{watanabe2025gabor3d,wurster2024gabor,zhou20253dgabsplat}, but these are designed for non-negative real-valued signals and do not directly extend to the complex-valued MRI setting.

In this work, we instead modulate each Gaussian with a complex exponential, yielding a freely positionable spectral component per primitive that can capture high-frequency content directly rather than through superposition of many narrow Gaussians. The formulation generalizes standard Gaussian primitives, which are recovered when the modulation frequency is zero, so low-frequency representation is preserved. We further propose a two-component low-rank temporal model that separates geometry dynamics from signal-intensity variations, and validate the framework on cardiac cine MRI with both Cartesian and non-Cartesian trajectories. Our contributions are:
\begin{enumerate}
\item We introduce a novel complex-valued Gabor primitive formulation for MRI reconstruction, where each primitive carries a freely positionable \textit{k}-space spectral component. This generalizes standard Gaussian primitives and enables efficient representation of both smooth anatomy and sharp boundaries.
\item We design a two-component low-rank temporal model that splits per-primitive dynamics into geometry and signal-intensity bases, providing a structured and compact parametrization of cardiac motion and contrast changes.
\item Across multiple cardiac cine datasets with Cartesian and radial trajectories at high acceleration factors, we demonstrate consistent improvements over compressed sensing, Gaussian primitive, and hash-grid INR baselines in all evaluation settings.
\end{enumerate}

\section{Methods}
\begin{figure}[!t]
   \centering
   \includegraphics[width=\linewidth]{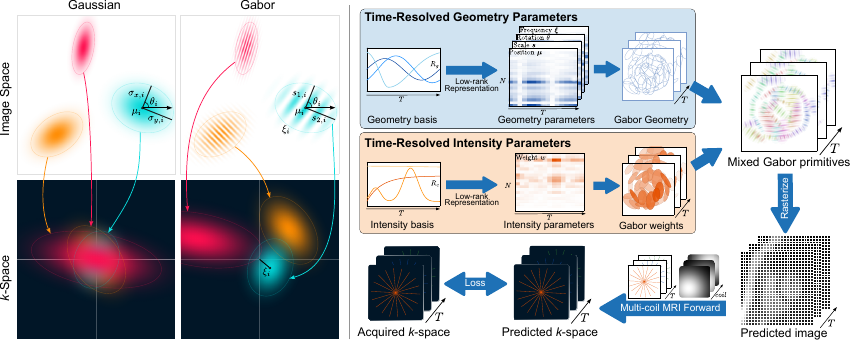}
   \caption{\textbf{Left:} Gaussian vs.\ Gabor primitives in image space and \textit{k}-space. A Gaussian's spectral support is fixed at the \textit{k}-space origin; a Gabor primitive shifts it to $\vec{\xi}_i$ via complex-exponential modulation, enabling more efficient frequency coverage. \textbf{Right:} Cardiac cine image is modeled as a mixture of time-varying Gabor primitives. Geometry parameters ($\vec{\mu}$, $\vec{s}$, $\theta$, $\vec{\xi}$) and complex weights $w$ are generated from low-rank geometry (blue) and intensity (orange) bases. Primitives are rasterized, passed through a multi-coil forward model, and fitted to acquired \textit{k}-space data end-to-end.}
   \label{fig:method}
\end{figure}

\subsection{Gabor Primitive Basis}
A standard Gaussian primitive centered at $\vec{\mu}_n$ with covariance $\Sigma_n$ defines a smooth spatial basis function. 
Its Fourier transform has a Gaussian envelope centered at the \textit{k}-space origin regardless of the primitive's spatial position, so representing high-frequency content requires superposition of many narrow primitives.
To overcome this limitation, we modulate each Gaussian with a complex exponential to form a Gabor primitive~\cite{daugman1985uncertainty,gabor1946theory,wurster2024gabor}:
\begin{equation}
P_n(\vec{r}) = \underbrace{\exp\!\Bigl(i\,2\pi\vec{\xi}_n \cdot (\vec{r} - \vec{\mu}_n)\Bigr)}_{\text{complex exponential}} \cdot \underbrace{\exp\!\Bigl(-\frac{1}{2}(\vec{r} - \vec{\mu}_n)^\top \Sigma_n^{-1} (\vec{r} - \vec{\mu}_n)\Bigr)}_{\text{Gaussian envelope}},
\label{eq:gabor}
\end{equation}
where $\vec{r} \in \mathbb{R}^2$ is the spatial coordinate, $\vec{\mu}_n \in \mathbb{R}^2$ is the center position, $\vec{\xi}_n \in \mathbb{R}^2$ is the modulation frequency, and $\Sigma_n = R_{\theta_n}\, \mathrm{diag}(s_{n,1}^2,\, s_{n,2}^2)\, R_{\theta_n}^\top$ is the covariance matrix parameterized by rotation angle $\theta_n$ and two scale parameters $s_{n,k} > 0$ along each principal axis. The Fourier transform of $P_n$ is:
\begin{equation}
\hat{P}_n(\vec{k}) = \frac{2\pi}{\sqrt{|\Sigma_n^{-1}|}}\, \exp\!\Biggl(-2\pi^2\bigl(\vec{k} - \vec{\xi}_n\bigr)^\top \Sigma_n \bigl(\vec{k} - \vec{\xi}_n\bigr)\Biggr)\, e^{-i2\pi\vec{k}\cdot\vec{\mu}_n},
\label{eq:gabor_ft}
\end{equation}
where $\vec{k} \in \mathbb{R}^2$ is the \textit{k}-space coordinate. Thus, each primitive produces a single Gaussian blob in \textit{k}-space centered at the modulation frequency $\vec{\xi}_n$ rather than at the origin (Fig.~\ref{fig:method}, left), reducing to a standard Gaussian when $\vec{\xi}_n = \vec{0}$. By the Fourier scaling property, the \textit{k}-space blob inherits the same orientation but with reciprocal widths. 
Compared to standard Gaussians whose spectra are all anchored at the \textit{k}-space origin, Gabor primitives place spectral support at arbitrary \textit{k}-space locations, reducing spectral overlap and improving representation efficiency.

\subsection{Spatiotemporal Forward Model}
\label{sec:model}
We model the cardiac cine MR image across $T$ frames as a sum of $N$ Gabor primitives whose weights and geometry may vary across frames:
\begin{equation}
x_t(\vec{r}) = \sum_{n=1}^{N} w_{n,t} \, P_n\!\bigl(\vec{r};\; \vec{\mu}_{n,t},\, \vec{s}_{n,t}, \theta_{n,t}, \vec{\xi}_{n,t}\bigr), \quad t = 1, \dots, T,
\label{eq:image_model}
\end{equation}
where $w_{n,t} \in \mathbb{C}$ is the complex weight. 
We model this temporal variation through two low-rank temporal bases (Fig.~\ref{fig:method}, right): a \emph{geometry basis} capturing cardiac motion and a \emph{intensity basis} modeling signal-intensity variations.
\subsubsection{Geometry Basis.}
Each geometric parameter is decomposed into a static component plus a low-rank perturbation from a shared basis $\mathbf{V}_g \in \mathbb{R}^{T \times R_g}$:
\begin{equation}
\begin{aligned}
\vec{\mu}_{n,t} &= \vec{\mu}_n + \mathbf{C}_n^{[\mu]}\vec{v}_{g,t},\quad &
\theta_{n,t} &= \theta_n + \vec{c}_n^{[\theta]} \cdot \vec{v}_{g,t},\\
\log\vec{s}_{n,t} &= \log\vec{s}_n + \mathbf{C}_n^{[s]}\vec{v}_{g,t},\quad &
\vec{\xi}_{n,t} &= \vec{\xi}_n + \mathbf{C}_n^{[\xi]}\vec{v}_{g,t},
\end{aligned}
\label{eq:geometry}
\end{equation}
where $\vec{v}_{g,t} \in \mathbb{R}^{R_g}$ is the $t$-th row of $\mathbf{V}_g$, $\log$ denotes the element-wise logarithm, 
and the $\mathbf{C}_n^{[\cdot]}$ are per-primitive coefficient matrices that determine how each primitive couples to the shared motion basis.
\subsubsection{Contrast Basis.}
In cardiac cine MRI, through-plane motion causes anatomical structures to move in and out of the imaging plane, producing frame-to-frame intensity variations beyond what in-plane geometry changes can explain. The weight $w_{n,t}$ captures these variations:
\begin{equation}
w_{n,t} = \vec{u}_n \cdot \vec{v}_{c,t} + \vec{c}_n^{[w]} \cdot \vec{v}_{g,t},
\label{eq:weights}
\end{equation}
where $\vec{v}_{c,t} \in \mathbb{C}^{R_c}$ is the $t$-th row of an independent intensity basis $\mathbf{V}_c \in \mathbb{C}^{T \times R_c}$, $\vec{u}_n \in \mathbb{C}^{R_c}$ are per-primitive intensity coefficients, and $\vec{c}_n^{[w]} \in \mathbb{C}^{R_g}$ couples the weight to the geometry basis to model motion-induced intensity changes such as partial-volume and flow effects. The resulting weight matrix $\mathbf{W} \in \mathbb{C}^{N \times T}$ has rank at most $R_c + R_g$, imposing a low-rank temporal structure directly in primitive parameter space.

\subsubsection{Multi-Coil Forward Model.}
Multi-coil \textit{k}-space data can be derived from the predicted images $x_t$ by applying the forward model for coil $j$ at frame $t$:
\begin{equation}
\hat{y}_{j,t} = \mathcal{F}_\Omega\!\bigl\{\mathcal{S}_j \, x_t\bigr\},
\label{eq:forward}
\end{equation}
where $\mathcal{S}_j$ denotes a pre-estimated coil sensitivity map
and $\mathcal{F}_\Omega$ is the Fourier transform evaluated on the acquired trajectory $\Omega$, implemented as a masked FFT for Cartesian sampling or a non-uniform FFT (NUFFT) for non-Cartesian trajectories. In practice, we first rasterize Eq.~\eqref{eq:image_model} on a Cartesian grid using the frame-specific geometry from Eq.~\eqref{eq:geometry} and Eq.~\eqref{eq:weights}, multiply the results by each coil sensitivity, and apply $\mathcal{F}_\Omega$ (Eq.~\eqref{eq:forward}). Overall, the model is optimized using the composite loss:
\begin{equation}
\mathcal{L} = \sum_{j=1}^{N_c} \sum_{t=1}^{T} \left\| y_{j,t} - \hat{y}_{j,t} \right\|_2^2
+ \lambda_s \sum_{n,t} |w_{n,t}|
+ \lambda_t \sum_{t=1}^{T-1} \left\| x_{t+1} - x_t \right\|_1,
\label{eq:loss}
\end{equation}
combining data fidelity between the predicted and measured \textit{k}-space, $\ell_1$ sparsity on primitive weights $w_{n,t}$ ($\lambda_s$), and temporal total variation ($\lambda_t$). 
All learnable quantities, including static primitive parameters, coefficient matrices, and temporal bases, are optimized jointly end-to-end.

\subsection{Training and Inference}
Primitive centers $\vec{\mu}_n$ are initialized on a uniform grid covering the imaging domain; modulation frequencies $\vec{\xi}_n$ are initialized near zero and evolve freely during optimization. Temporal coefficient matrices are initialized to zero, meaning that the model starts static and gradually discovers motion. We adopt adaptive density control~\cite{kerbl20233dgs}, periodically pruning low-contribution primitives and splitting those with large gradients. Optimization uses Adam~\cite{kingma2014adam} with per-parameter-group learning rates and cosine annealing. At inference, the continuous representation can be evaluated on any spatial grid.

\section{Experiments}

\subsubsection{Datasets.}
We evaluate on two cardiac cine MRI datasets.
\textbf{(1) Cartesian:} 99 fully sampled bSSFP cardiac cine acquisitions (55 at 1.5\,T, 44 at 3\,T, $208{\times}256$, ${\approx}19$ frames, 15--18 coils) from~\cite{chen2020ocmr}, retrospectively undersampled at $R{=}12$ and $R{=}16$ with VISTA masks~\cite{ahmad2015variable}. Coil sensitivities were estimated via ESPIRiT~\cite{uecker2014espirit}.
\textbf{(2) Radial:} 102 radial bSSFP acquisitions at 3\,T from~\cite{elrewaidy2020harvard} ($208{\times}208$, 25 cardiac phases, coils compressed to 8), retrospectively subsampled to 14 spokes per frame ($R{\approx}23$). References were obtained via CG-SENSE from all spokes.

\subsubsection{Baselines.}
We compare against two conventional methods and two scan-specific learned representations: (1)~\textbf{L+S}~\cite{otazo2015low}, robust PCA decomposition into low-rank plus sparse components; (2)~\textbf{PICS}~\cite{lustig2007sparse}, iterative parallel imaging and compressed sensing with total variation regularization; (3)~\textbf{Hash-INR}, a scan-specific implicit neural representation with multi-resolution hash-grid encoding~\cite{muller2022instant}; and (4)~\textbf{Gaussian primitives}, our framework with $\vec{\xi}_n \equiv \vec{0}$, serving as an ablation baseline.

\subsubsection{Implementation Details.}
All methods were implemented in PyTorch on an NVIDIA A6000 GPU, with custom CUDA kernels for primitive rasterization. To match total parameter counts, Gabor used $N{=}15{,}000$ initial primitives growing to $20{,}000$, while Gaussian used $19{,}000$ growing to $21{,}000$. Both used temporal ranks $R_g{=}6$, $R_c{=}4$ with geometry--contrast coupling, adaptive density control (every 300 iterations, 10--60\% of training), and $5{,}000$ iterations ($\lambda_s{=}10^{-5}$, $\lambda_t{=}10^{-2}$).
Hash-INR used 16 hash levels ($2^{19}$ map size), a 2-hidden-layer MLP (128 nodes) and trained for $5{,}000$ iterations.
Quality was evaluated by peak signal-to-noise ratio (PSNR), structural similarity (SSIM), and feature similarity (FSIM)~\cite{zhang2011fsim}.
Temporal ranks and regularization weights were selected via grid search on a representative subset of five scans.

\section{Results}

\subsubsection{Quantitative Comparison.}
Table~\ref{tab:main} summarizes reconstruction quality across all test slices.
\begin{table}[t]
\caption{Quantitative comparison on cardiac cine MRI ($N{=}99$ Cartesian, $N{=}102$ radial). PSNR in dB, Time in minutes. $\rho = N_{\text{params}} / (2HWT)$: ratio of learnable parameters to real-valued image degrees of freedom. Best in \textbf{bold}, second best \underline{underlined}.}\label{tab:main}
\centering
\resizebox{\textwidth}{!}{%
\begin{tabular}{lc|cccc|cccc|cccc}
\toprule
 & & \multicolumn{4}{c|}{Cartesian $R{=}12$} & \multicolumn{4}{c|}{Cartesian $R{=}16$} & \multicolumn{4}{c}{Radial $R{\approx}23$} \\
Method & $\rho\downarrow$ & PSNR$\uparrow$ & SSIM$\uparrow$ & FSIM$\uparrow$ & Time$\downarrow$ & PSNR$\uparrow$ & SSIM$\uparrow$ & FSIM$\uparrow$ & Time$\downarrow$ & PSNR$\uparrow$ & SSIM$\uparrow$ & FSIM$\uparrow$ & Time$\downarrow$ \\
\midrule
L+S~\cite{otazo2015low}               & --- & 37.15 & 0.940 & 0.850 & \underline{0.3} & 32.19 & 0.824 & 0.777 & \underline{0.4} & 35.82 & 0.922 & 0.810 & \underline{1.5} \\
PICS~\cite{lustig2007sparse}             & --- & \underline{42.21} & \underline{0.972} & \textbf{0.886} & \textbf{0.1} & \underline{41.05} & 0.966 & \textbf{0.875} & \textbf{0.1} & 35.19 & 0.905 & 0.793 & \textbf{1.1} \\
Hash-INR~\cite{muller2022instant} & 2.60 & 39.63 & 0.921 & 0.841 & 4.6 & 37.77 & 0.893 & 0.815 & 4.6 & 34.00 & 0.849 & 0.767 & 8.0 \\
Gaussian       & \underline{0.42} & 41.50 & 0.972 & 0.874 & 1.2 & 40.67 & \underline{0.967} & 0.864 & 1.2 & \underline{36.67} & \underline{0.942} & \underline{0.810} & 3.4 \\
\textbf{Gabor (ours)}                  & \textbf{0.41} & \textbf{42.61} & \textbf{0.974} & \underline{0.883} & 2.5 & \textbf{41.39} & \textbf{0.968} & \underline{0.871} & 2.5 & \textbf{37.53} & \textbf{0.947} & \textbf{0.830} & 4.2 \\
\bottomrule
\end{tabular}}
\end{table}
Gabor primitives achieve the highest PSNR and SSIM across all three settings, with consistent gains over Gaussian ($+$1.11/$+$0.72/$+$0.86~dB). 
While PICS achieves marginally higher FSIM on Cartesian data, Gabor primitives outperform all competing methods on radial data ($R{\approx}23$), including a $+$2.34~dB gain over PICS. This highlights the stronger implicit structural prior of Gabors compared to TV regularization under severe undersampling, while offering the most compact image representation ($\rho{<}0.5$).
Hash-INR requires $6{\times}$ more parameters ($\rho{=}2.60$) yet ranks last among learned methods. Gabor primitives are slower than Gaussian because they capture high frequencies via modulation rather than spatial narrowing, resulting in spatially wider primitives on average. During rasterization, each pixel therefore overlaps with more Gabor than Gaussian primitives, increasing per-pixel computation.

\subsubsection{Qualitative Comparison.}
\begin{figure}[t]
   \centering
   \includegraphics[width=\textwidth]{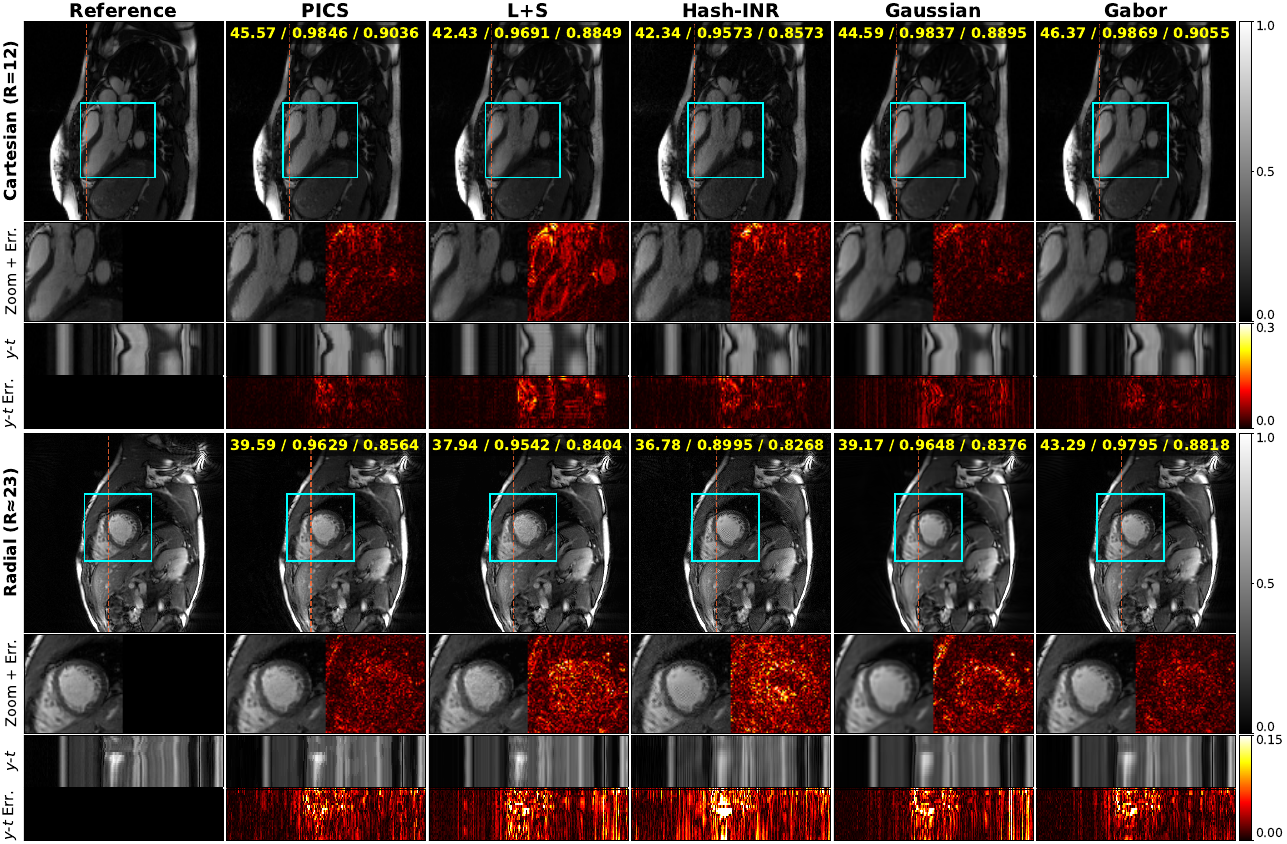}
   \caption{Reconstruction comparison on Cartesian (top) and radial (bottom) cardiac cine. Each block shows a representative frame with metrics, zoomed ROI with error map, and $y$--$t$ profile with temporal error.}
   \label{fig:recon}
\end{figure}

Figure~\ref{fig:recon} compares representative reconstructions across all methods.
Voxel-based methods (PICS, L+S, Hash-INR) show visible background noise in the error maps, since their per-voxel degrees of freedom allow fitting noise alongside signal. Primitive methods suppress this via locally smooth, compact representations ($\rho{<}0.5$). L+S blurs temporal motion boundaries in the $y$--$t$ profiles, and all non-Gabor methods degrade more on radial data. Gaussian shows elevated error at tissue boundaries under high acceleration (radial zoom), where the absence of carrier frequency $\vec{\xi}$ limits edge representation. Gabor yields the lowest spatial and temporal error across both settings.

\subsubsection{Representation Properties.}
Beyond reconstruction quality, Gabor primitives provide unique capabilities over grid-based and Gaussian methods.

\begin{figure}[t]
   \centering
   \includegraphics[width=\textwidth]{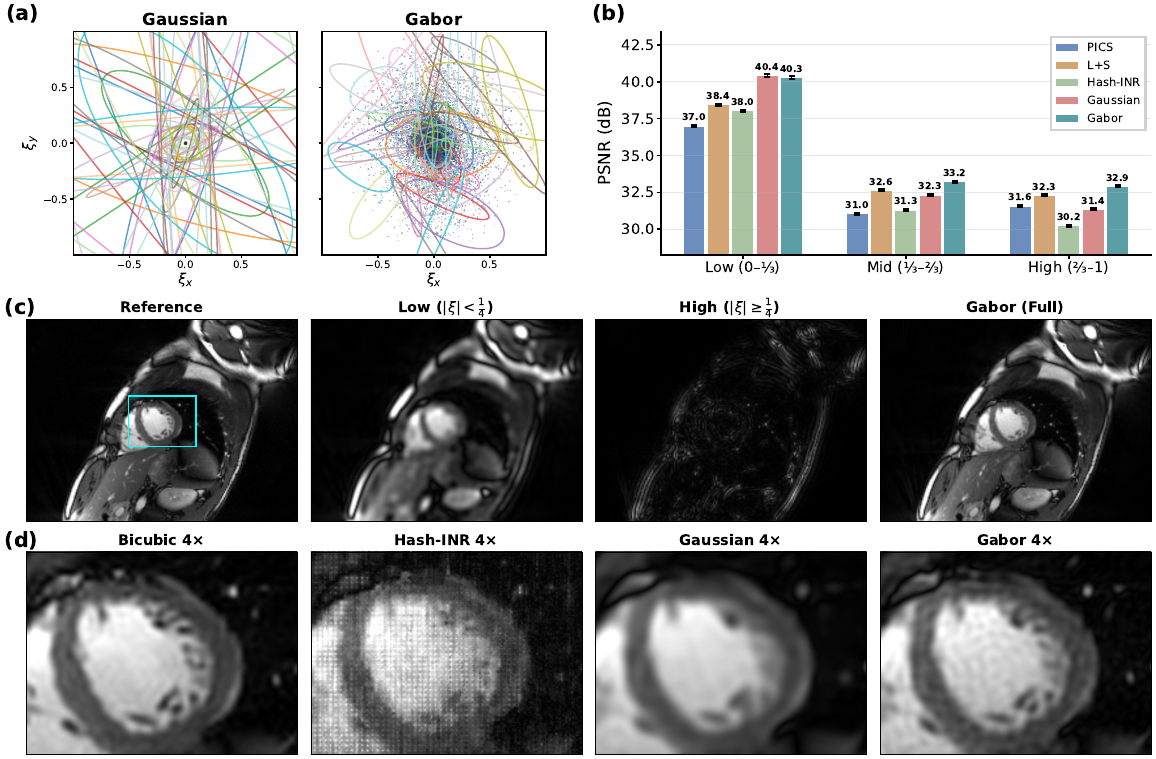}
   \caption{Unique capabilities of Gabor primitives. \textbf{(a)}~\textit{k}-space primitive distribution: each dot is a primitive's center frequency $\vec{\xi}_n$ with $3\sigma$ support ellipses. 
   \textbf{(b)}~Frequency-band PSNR (low, mid, high) on the radial dataset, averaged over all subjects. \textbf{(c)}~Frequency decomposition: primitives partitioned by $|\vec{\xi}_n|$ into low- ($<\!\tfrac{1}{4}$) and high-frequency ($\geq\!\tfrac{1}{4}$) groups and their sum. \textbf{(d)}~$4\times$ super-resolution on the marked ROI; bicubic upsampling of the reference shown for comparison.}

   \label{fig:unique}
\end{figure}

\textbf{\textit{k}-space coverage.} Figure~\ref{fig:unique}a shows that learned Gabor primitives distribute their center frequencies $\vec{\xi}_n$ across \textit{k}-space to match the image's spectral content; the overlaid $3\sigma$ support ellipses illustrate the anisotropic spectral footprint of individual primitives. In contrast, Gaussian primitives are anchored at the origin. This spectral advantage is confirmed quantitatively in Fig.~\ref{fig:unique}b: across the radial cohort, all methods achieve similar PSNR in the low-frequency band, while Gabor yields the largest gains in the mid- and high-frequency bands where Gaussian and L+S fall short.

\textbf{Frequency decomposition.} Partitioning primitives by $|\vec{\xi}_n|$ yields a natural spectral decomposition (Fig.~\ref{fig:unique}c): low-frequency primitives capture smooth anatomy, while high-frequency ones capture edges (no Gaussian counterpart, where $|\vec{\xi}_n|=0$).

\textbf{Super-resolution.} As a continuous representation, Gabor primitives can be evaluated at arbitrary resolution without retraining (Fig.~\ref{fig:unique}d). We compare against a bicubic-upsampled as reference. Gabor primitives recover sharper structures than Gaussian primitives, which lack the carrier frequency $\vec{\xi}$ needed to extrapolate high-frequency detail. Hash-INR does not inherently guarantee spatial continuity and, due to overfitting on the sampled grid, produces visible raster artifacts at the increased resolution.

\section{Discussion and Conclusion}

We proposed Gabor primitives for MRI reconstruction, where complex exponential modulation gives each primitive a freely positionable \textit{k}-space spectral component, enabling efficient high-frequency coverage that Gaussian primitives cannot achieve. A two-component temporal model factorizes per-primitive dynamics into geometry and intensity bases, imposing a structured low-rank prior in primitive parameter space.
Across multiple cardiac cine settings with Cartesian and radial trajectories, Gabor primitives consistently outperform all baselines. The explicit parameterization also offers physical interpretability, such as spectral decomposition by modulation frequency, and could provide compact descriptors for downstream tasks such as motion quantification in the future.

Several limitations remain. As a scan-specific method, each acquisition requires individual optimization (2--4 minutes). The formulation operates in 2D; extending to 3D and joint spatiotemporal frequency modulation could improve temporal modeling. Validation on additional anatomies and clinical diagnostic evaluation are needed for translation.

\section*{Acknowledgements}
This work was partially funded by the European Research Council (ERC) under Grant Deep4MI (No. 884622). The authors acknowledge GE HealthCare, Munich, for providing data for early-stage testing in this study.
V.S. was supported by the Add-On Fellowship of the Joachim Herz Foundation.
Dr. Sevgi Gokce Kafali has been sponsored by the Alexander von Humboldt Foundation.
SK was supported by the National Institute of Diabetes and Digestive and Kidney Diseases (NIDDK) of the National Institutes of Health under Award No. R01DK125561 and by the Massachusetts AI Hub Award.

%
%
\bibliographystyle{splncs04}
\bibliography{references}

\end{document}